\documentclass[twocolumn,amsmath,amssymb,showpacs]{revtex4}

\usepackage{color}
\usepackage{amsmath}
\usepackage{graphicx}
\usepackage{makecell}

\begin{document}
\title{The stability and properties of high-buckled two-dimensional tin and lead}
\author{Pablo Rivero,$^1$ Jia-An Yan,$^2$ V{\'\i}ctor M. Garc{\'\i}a-Su{\'a}rez,$^3$ Jaime Ferrer,$^3$ and Salvador Barraza-Lopez$^1$}
\email{sbarraza@uark.edu}
\affiliation{1. Department of Physics. University of Arkansas. Fayetteville AR, 72701. USA.\\
2. Department of Physics, Astronomy and Geosciences. Towson University. Towson, MD 21252 USA.\\
3. Departamento de F{\'\i}sica and Centro de Investigaci{\'o}n en Nanociencia y Nanotecnolog{\'\i}a. Universidad de Oviedo.}
\begin{abstract}
 In realizing practical non-trivial topological electronic phases stable structures need to be determined first. Tin and lead do stabilize an optimal two-dimensional high-buckled phase --a hexagonal-close packed
bilayer structure with nine-fold atomic coordination-- and they do not stabilize topological fullerenes, as
demonstrated by energetics, phonon dispersion curves, and the structural optimization of finite-size
samples. The high-buckled phases are metallic due to their high atomic coordination. The optimal structure
of fluorinated tin lacks three-fold symmetry and it stabilizes small samples too. It develops two oblate conical valleys on the first Brillouin
zone coupling valley, sublattice, and spin degrees of freedom with a novel $\tau_z\sigma_xs_x$ term, thus making it a new 2D
platform for valleytronics.
\end{abstract}
\date{\today}

\maketitle
\noindent{\em Introduction.- }Carbon forms two-dimensional (2D) layers with a hexagonal lattice \cite{KatsnelsonBook,RMP} and
silicon, germanium \cite{Ciraci2009}, AlAs, AlSb, GaP, InP, GaAs, InAs, GaSb, InSb \cite{Hennig2013}, phosphorus \cite{Tomanek2014},
and tin \cite{jpc1,jpc2,stanene} are all predicted to form stable low-buckled (LB) hexagonal 2D layers. High-buckled (HB) 2D phases
cannot occur for carbon, silicon, nor germanium \cite{Ciraci2009,AMI}. Can 
 tin and lead stabilize the HB phase?

Proceeding by direct analogy to silicene and germanene~\cite{Ciraci2009}, known studies of the electronic properties of 2D
tin \cite{jpc1,jpc2,stanene,2D} are performed under the implicit assumption  that the HB phase is not viable. In
addition, the guess structures and the electronic gaps in Ref.~\cite{stanene} had been previously reported \cite{jpc2}. Contrary
to common assumption, the HB 2D structures of heavy column-IV elements tin and lead
are stable and lower in energy than their LB counterparts, thus representing the true optimal structures of these two-dimensional systems.
 The structural stability of HB tin and HB lead will have fundamental consequences for the practical realization of substrate-free non-trivial
 topological phases based from these elements.

 The optimal phase of 2D fluorinated stanene is not analogous to tetrahedrally-coordinated
 graphane \cite{graphane} as it was postulated in Refs.~\cite{jpc2,stanene}. Studies of 2D fluorinated tin dismiss the existence of bulk crystalline
 fluorinated phases stable at room temperature. {\em There is no indication for
 tetrahedral coordination of tin atoms} in bulk fluorinated tin ~\cite{mcdonald} and tetrahedral coordination~\cite{jpc2,stanene} does not yield the
 most stable 2D fluorinated tin either.

\begin{figure}[tb]
\includegraphics[width=0.475\textwidth]{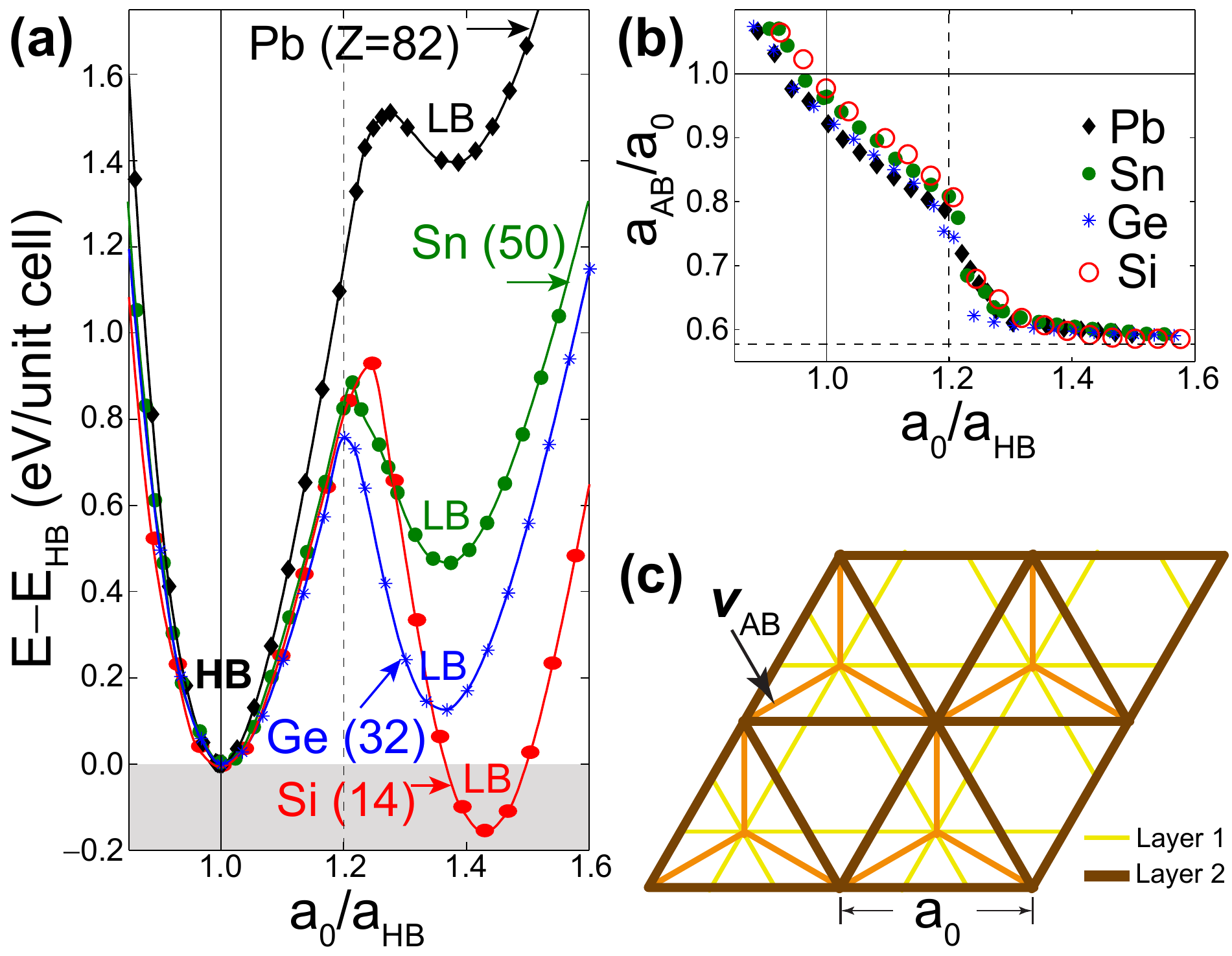}
\caption{(Color online.) (a) The high-buckled phase becomes more stable with increasing atomic number.
(b) Nearest-neighbor distances $a_{AB}\equiv |\mathbf{v}_{AB}|$ approach the lattice constant $a_0$ ($a_{AB}\simeq a_0$) at the high-buckled
energy minimum; the structure transitions to a low-buckled phase at roughly 1.2$a_{HB}$. (c) The high-buckled structure is a HCP bilayer.}\label{fig:F1}
\end{figure}

\begin{figure*}[tb]
\includegraphics[width=.95\textwidth]{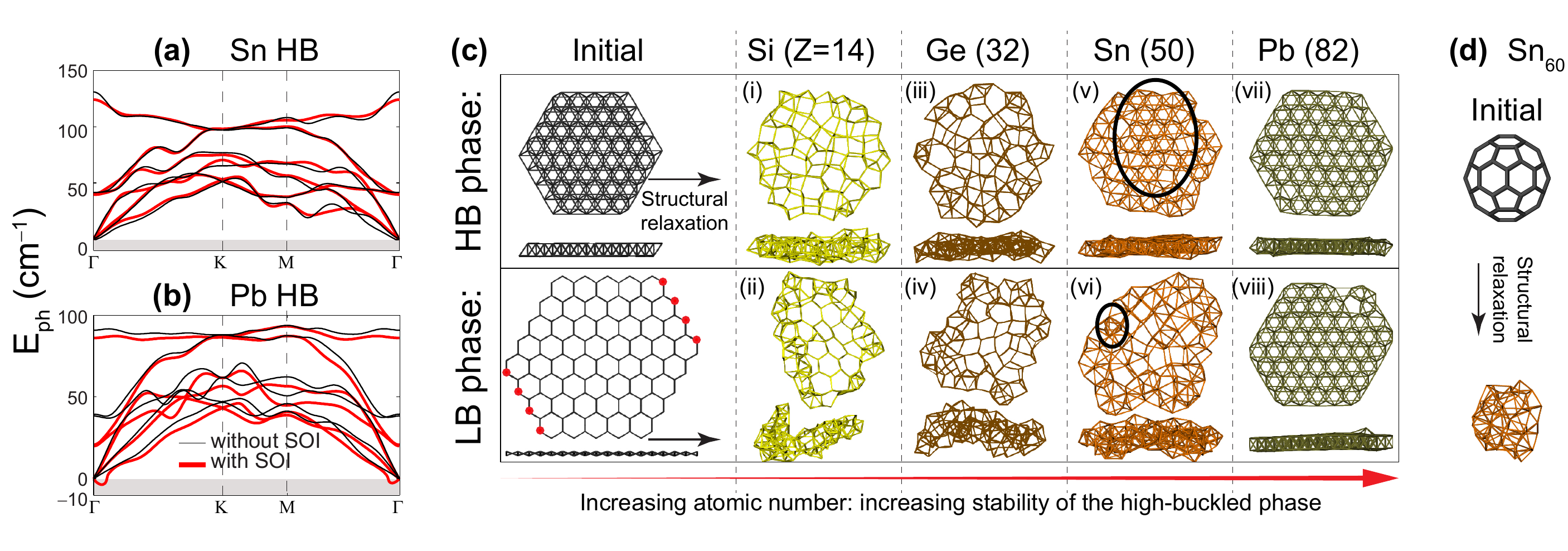}
\caption{(Color online.) (a-b) Phonon dispersions $E_{ph}(\mathbf{k})$ for HB tin and lead demonstrate their structural stability; SOI does not
change phonon dispersions dramatically. (c) The structural optimization of finite samples reflects previous findings for Si and Ge \cite{Ciraci2009}
and helps in confirming the stability of HB tin and HB lead unequivocally: The stanene sample (LB tin) becomes thick and amorphous and the initial
LB lead sample turns into HB lead. (d) 2D tin and lead do not realize topological fullerenes.}\label{fig:F2}
\end{figure*}

  We uncover {\em six metastable fluorinated phases for 2D tin}, the graphane-like phase \cite{jpc1,jpc2,stanene} being one of them.
 Consistent with the literature in bulk flourinated tin \cite{mcdonald,denes}, we demonstrate that {\em two tilted F atoms mediate
 the interaction among two Sn atoms} in the optimal 2D structure. This stable optimal phase displays two gapped oblate Dirac cones on the first Brillouin zone where valley
 $\boldsymbol{\tau}$, pseudospin $\boldsymbol{\sigma}$, and spin $\mathbf{s}$ couple as $\tau_z\sigma_xs_x$ \cite{KaneMele}.

 Unlike known 2D materials with a hexagonal lattice in which three valleys with momentum directions
 separated by 120$^o$ rotations are related due to threefold symmetry \cite{beenaker,Xiao,McDonald2,Guinea,NatPhys14}, the optimal
 2D flourinated tin leads to {\em strictly two} valleys due to its reduced structural symmetry. This allows an unprecedented specificity in
 coupling three quantum degrees of freedom around the Fermi energy: the valley, the crystal momentum {\em including direction},
and the electronic spin.

 The results here provided invite to look closely into 2D materials postulated for their
 remarkable electronic properties that may not realize ground-state, optimal structures \cite{jpc2,stanene}.

\noindent{\em The HB phase is more favorable than the LB phase with increasing atomic number.- } The energetics of column-IV 2D
materials on Fig.~1(a) were obtained with the PBE exchange-correlation potential \cite{PBE} on a version of the {\em SIESTA} code \cite{SIESTA1,SIESTA2}
that includes a self-consistent spin-orbit interaction (SOI) \cite{Lucas}. Our basis sets are of double-zeta plus-polarization size \cite{Pablo1}.  The
trends in Fig.~1 remain regardless of the inclusion of SOI, and were cross-checked with {\em VASP} calculations \cite{VASP1,VASP2}.

 The lattice constant at the HB energy minima $a_{HB}$ is equal to 3.418  \AA{} for tin, and $a_{HB}$=3.604 \AA{} for lead. These values become
 3.413 and 3.575 \AA, respectively, when the SOI is included in calculations. These strikingly stable HB structures have not been reported
 before; lattice parameters in the literature \cite{jpc1,jpc2,stanene,2D} are $\sim$140\% larger. Normalization of $a_0$ in terms of $a_{HB}$ in
 Figs.~1(a) facilitates an unified display of energetics regardless of atomic species.  The vertical dashed line in Figs.~1(a) and 1(b) at about
 $a_0 \simeq 1.2 a_{HB}$ highlights the lattice constant $a_0$ for which energy barriers separating the LB and the HB phases become largest.

 Germanium (with atomic number $Z=32$) cannot form a HB phase, even though the energy minima of the optimized HB phase is lower than the local
 minima at the optimal LB phase already [Fig.~1(a)] \cite{Ciraci2009}. This HB minima becomes sizeable deeper and the energy barriers separating
 these phases become shallower with increasing atomic number. Figure 1(a) invites to ponder whether HB tin and HB lead are stable. In answering
 this question we address the atomistic coordination of HB phases first.

\noindent{\em The optimal HB structure is a hexagonal close-packed bilayer.- } HB phases were represented as three-fold coordinated \cite{Ciraci2009},
but the relative height $\Delta z$ among atoms in complementary sublattices $A$ and $B$ increases as the lattice constant $a_0$ is compressed, so the
distance $a_{AB}=\sqrt{a_0^2/3+\Delta z^2}$ among atoms belonging in complementary sublattices increases towards $a_0$. Indeed, $a_{AB}=a_0/\sqrt{3}$
for a planar hexagonal unit cell --dashed horizontal line on Fig.~1(b)-- but an ideal hexagonal close-packed (HCP) structure has
$\Delta z=\sqrt{2}a_0/\sqrt{3}$ yielding $a_{AB}=a_0$  --solid horizontal line on Fig.~1(b) \cite{Marder}. Numerical results yield $a_{AB}\simeq 0.95a_{HB}$
(solid vertical line on Fig.~1(b)). Thus, six atoms are a distance $a_{HB}$ apart on a triangular lattice, and three atoms belonging on complementary
sublattices are separated by $a_{AB}\simeq0.95 a_{HB}$, leading to the nine-fold coordinated HCP bilayer structure \cite{Palacios,newer}
on Fig.~1(c). A transition  among low- and high-buckled structures occurs around $a_0 \simeq 1.2 a_{HB}$ on Fig.~1(b).

\noindent{\em HB tin and HB lead are stable.- } We show in Figs.~2(a-b) phonon dispersion curves for HB tin and lead \cite{note1}. The effect of SOI
is small, thus justifying the trends without SOI shown on Fig.~1(a-b) \cite{note4}. Similar dispersions were obtained using the {\em Quantum Espresso}
code \cite{expresso}. The
lack of significant negative energies indicates that HB tin and HB lead are indeed stable: The
Chemistry of Si and Ge does not translate to Sn and Pb because with increasing atomic number the $s-$orbital lowers its energy
with respect to $p-$orbital, thus reducing the $s-p$ hybridization.

The ultimate test of relative stability is a structural optimization of small 2D flakes with initial HB or LB conformations [Fig.~2(c)] where the lines
joining atoms reveal their atomistic coordination. The finite-size HB structures have 122 atoms; the LB structures have eight additional atoms (red dots
on the LB initial structure) so that all edge atoms are two-fold coordinated. We set a stringent force tolerance cutoff of at least 0.01 eV/\AA{}.

   The LB Si and LB Ge samples (subplots $ii$ and $iv$) show crumpling originating out from the boundaries yet the hexagonal lattice remains visible
   around the center of mass after the force relaxation \cite{Ciraci2009}. On the other hand, the amorphous shape and the random-looking atomistic
   coordination of HB Si and HB Ge (subplots $i$ and $iii$) indicate that these phases are unstable \cite{Ciraci2009}.

 Confirming the structural stability inferred from phonon dispersion curves, HB Sn and HB Pb do stabilize on finite-size samples: Starting from an ideal
 HB phase, the optimized Sn structure retains the HB coordination within the area highlighted by an oval (Fig.~2(c), subplot $v$). The finite LB Sn sample,
 on the other hand, crumples upon optimization (Fig.~2(c), subplot $vi$). In fact, the region highlighted by the tiny oval on subplot $vi$ in Fig.~2(c)
 displays the local coordination expected of a HB phase already. Similar conclusions would be reached in Ref.~\cite{2D} when periodic constraints
 are removed.

 Haldane's honeycomb model has been studied in closed geometries \cite{circular} and one of the many candidates for its practical realization is
 LB tin (stanene). Unfortunately, a fullerene-like Sn$_{60}$ is not stable [Fig.~2(d)] so tin and lead are no-go elements for topological fullerenes. Based
 on Fig.~1(a) HB lead is extremely stable: It stabilizes finite HB samples with no change in atomic coordination [Fig.~2(c), subplot $vii$] and turns
 a LB structure onto a HB-coordinated one [Fig.~2(c), subplot $viii$].

  Viable electronic materials require
  stable structures. Tin and lead films have been created experimentally \cite{films1,films2,my1} and structural aspects must be addressed
  diligently to realize two-dimensional materials with a strong SOI.

\noindent{\em Electronic properties of HB tin and lead.- } Graphene, silicene and germanene are three-fold coordinated and have a conical dispersion
around the $K-$points with small gaps due to SOI \cite{KaneMele,SOI1,SOI3,SOI4,SOI5}. The nine-fold-coordinated 2D HB structures display no conduction
gaps [Fig.~3]. 

\begin{figure}
\includegraphics[width=0.475\textwidth]{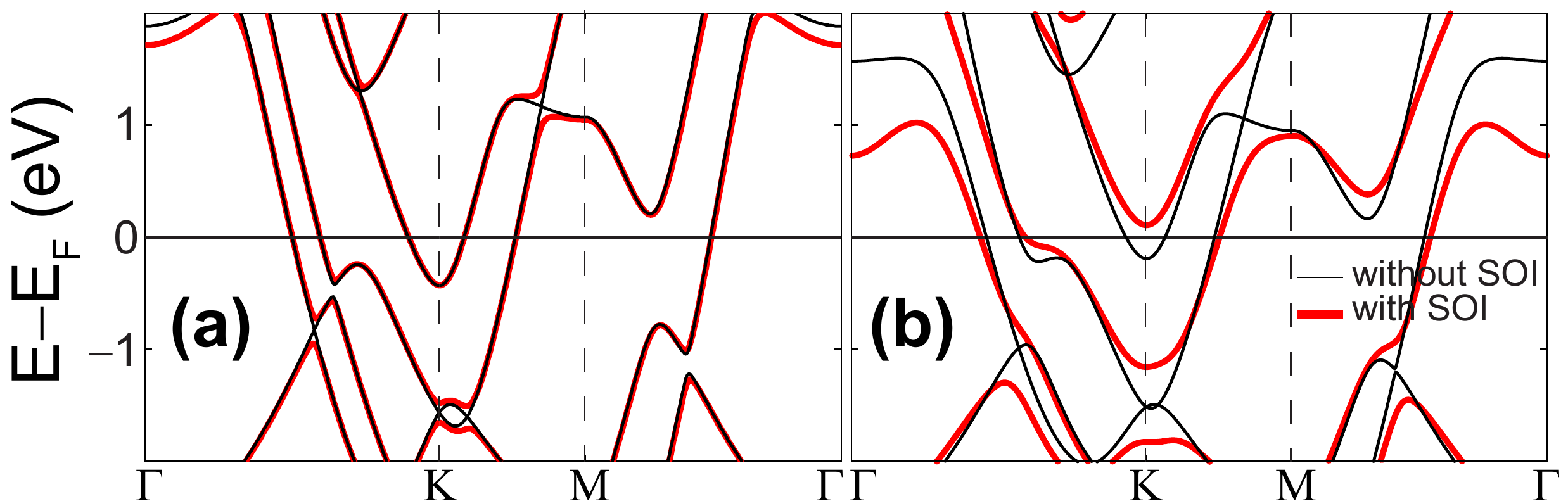}
\caption{(Color online.) Electronic dispersion for (a) HB tin and (b) HB lead. The nine-fold atomic coordination of the HB phases is behind the
metallic electronic dispersion.}\label{fig:F3}
\end{figure}

\noindent{\em Bulk limits.- } HCP bilayers could be cleaved out of HCP or FCC bulk structures. Lead forms a FCC structure with
interatomic distances  of 3.614 \AA{}, which compare favorably with $a_{HB}=3.575$ and make HB lead stable.

 Tin stabilizes a tetragonal structure ($\beta-$tin \cite{betatin,alphatin2}) and a diamond structure ($\alpha-$tin \cite{alphatin}). The $\beta-$phase
 is higher in energy than the $\alpha-$phase by $E_{\beta}-E_{\alpha}$= 0.58 eV/atom. Every atom on $\beta-$tin has four neighbors at 3.11 \AA, two
 neighbors at  $3.26$ \AA, and four neighbors 3.87 \AA{} apart: these  ten atoms are 3.44 \AA{} apart on average. On the nine-fold coordinated HB
 tin $a_{HB}=3.42$ \AA{} and $a_{AB}=3.281$, having an atomistic coordination comparable to bulk $\beta-$tin. The
 $\alpha-$tin phase has four neighbors  2.89 \AA{} apart, which compares well to a$_{Sn-Sn}=2.85$ \AA{} for a 2D LB structure.
Importantly, in two-dimensions energetics switch and the HB phase --compatible with bulk $\beta-$tin-- is more stable than LB tin --compatible with
$\alpha-$tin-- by $E_{LB}-E_{HB}=0.25$ eV/atom [c.f. Fig.~1(a)].

\begin{figure}[tb]
\includegraphics[width=.475\textwidth]{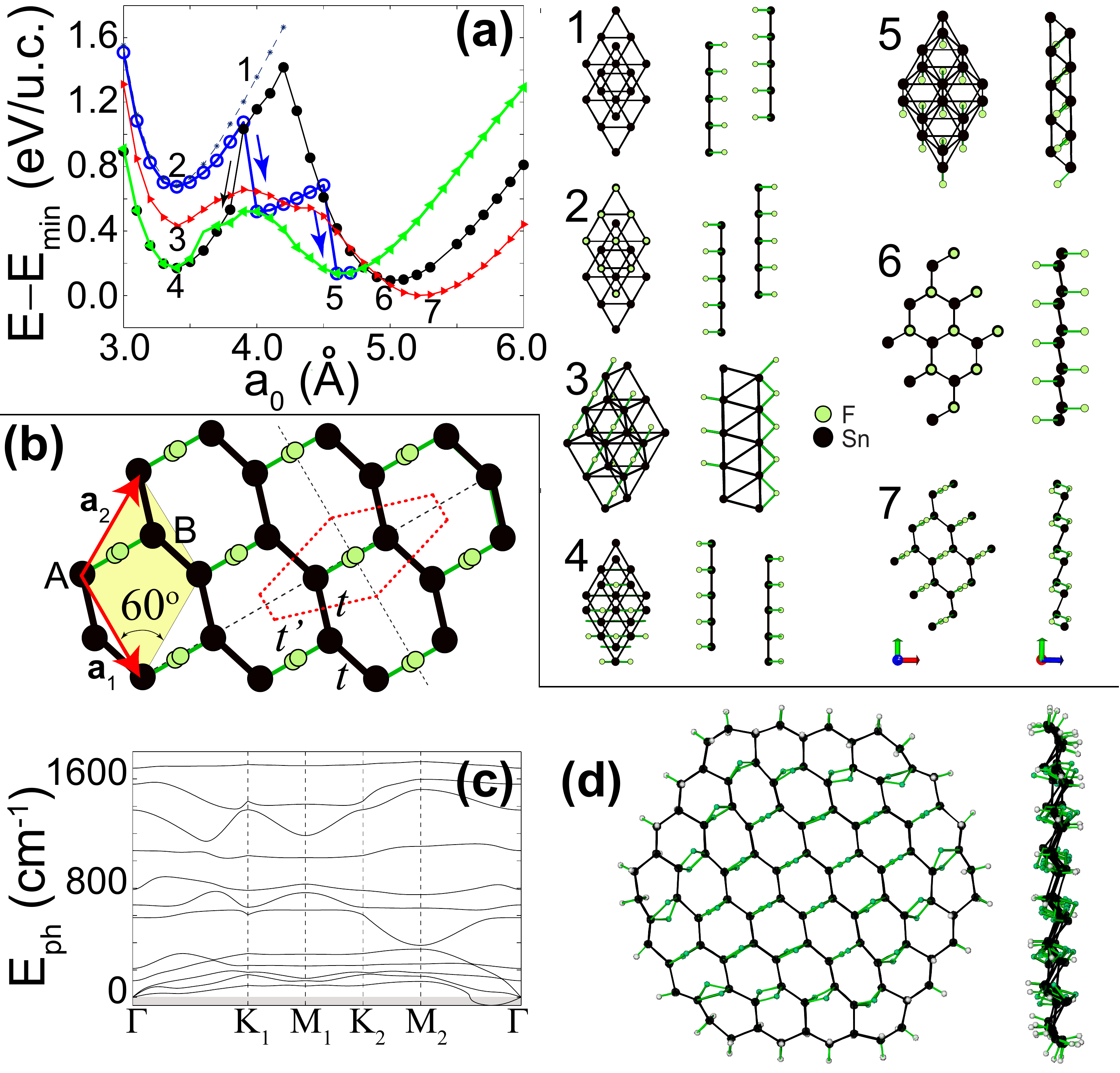}
\caption{(Color online.) (a) Phases of 2D fluorinated tin; structures shown to the right. (b) Symmetries of the most stable structure (7),
depicting
triangular (dashed) and Wigner-Seitz (within dotted perimeter) unit cells, the two symmetry axes, and the two Sn sublattices $A$ and $B$.
Structural
stability is demonstrated by (c) phonon dispersion curves and (d) the structural stabilization of a finite-size sample.}\label{fig:F4}
\end{figure}

\noindent{\em Fluorinated 2D tin.- } The phase space for decorated 2D tin is larger than originally anticipated [Fig.~4(a)]: The
graphane-like  phase \cite{jpc2,stanene} realizes the metastable minima labeled {\bf 6} that turns into phase {\bf 4} upon in-plane compression. Placement of
F atoms directly on top of/under Sn atoms results on two dissociated triangular Sn lattices bonded on opposite sides by F atoms (structures {\bf 2}/{\bf 1}).

  In the optimal structure, {\bf 7}, four-fold coordinated Sn atoms form a sequence of parallel zig-zag one-dimensional chains
  with two fluorine atoms mediating interactions among neighboring Sn chains.  The structure is realized on a triangular lattice with $a_0=5.230$ \AA{}
  [Fig.~4(b)]. The Wigner-Seitz unit cell is within the dotted area in Fig.~4(b); the symmetry axes are shown as well. A similar ``bridging'' fluorine coordination is realized on bulk tin(II) fluoride (e.g., Fig.~2 in Ref.~\cite{mcdonald}).

\begin{table}
\caption{Basis vectors for fluorinated stanene ($a_0=5.23$ \AA).}
\label{ta:ti1}
\begin{tabular}{ccc}
\hline\hline
Sn:&(0.000,	0.000,	0.000)$a_0$,& (0.583,	0.336,	 $-$0.221)$a_0$\\
\hline
F: &(0.216,	0.124,	$-$0.348)$a_0$,&(0.367,	0.212,	0.128)$a_0$\\
\hline\hline
\end{tabular}
\end{table}

 Bulk Tin(II) fluoride is highly stable at room temperature and
  can be found in household products. Structural stability of optimal 2D tin is probed with phonon dispersion calculations [Fig.~4(c)] along the high-symmetry lines
  shown in Fig.~5(a). The phonon frequency range is comparable with that of graphene, and it is one order-of-magnitude larger than those in Fig.~2(a,b). As an
  additional successful check, small-size flakes were subjected to a successful structural optimization [Fig~4(d)]. The peculiar coupling
  of quantum degrees of freedom on this system may encourage experimental routes towards the synthesis of 2D fluorinated tin. The stability of its parent
  3D compound at room temperature \cite{mcdonald,denes} invites experimental investigations of potential viability in 2D.

\begin{table}
\caption{Eigenvectors of $\tau_z\sigma_x s_x$. $|s_x;\pm\rangle$ are eigenstates of $s_x$, and $|A\rangle$, $|B\rangle$ are eigenstates of the pseudospin operator.}
\label{ta:ti3}
\begin{tabular}{ccc}
\hline\hline
State & $V_1$ & $V_2$\\
\hline
$|\phi_{-\Delta,1}\rangle$:&$\frac{1}{\sqrt{2}}(-|A\rangle+|B\rangle)|s_x;+\rangle$ &$\frac{1}{\sqrt{2}}(|A\rangle-|B\rangle) |s_x;-\rangle$\\
$|\phi_{-\Delta,2}\rangle$:&$\frac{1}{\sqrt{2}}(|A\rangle+|B\rangle) |s_x;-\rangle$ &$-\frac{1}{\sqrt{2}}(|A\rangle+|B\rangle) |s_x;+\rangle$\\
$|\phi_{+\Delta,1}\rangle$:&$\frac{1}{\sqrt{2}}(|A\rangle-|B\rangle) |s_x;-\rangle$ &$\frac{1}{\sqrt{2}}(-|A\rangle+|B\rangle) |s_x;+\rangle$\\
$|\phi_{+\Delta,2}\rangle$:&$\frac{1}{\sqrt{2}}(|A\rangle+|B\rangle) |s_x;+\rangle$ &$-\frac{1}{\sqrt{2}}(|A\rangle+|B\rangle) |s_x;-\rangle$\\
\hline\hline
\end{tabular}
\end{table}

\begin{figure}[tb]
\includegraphics[width=.475\textwidth]{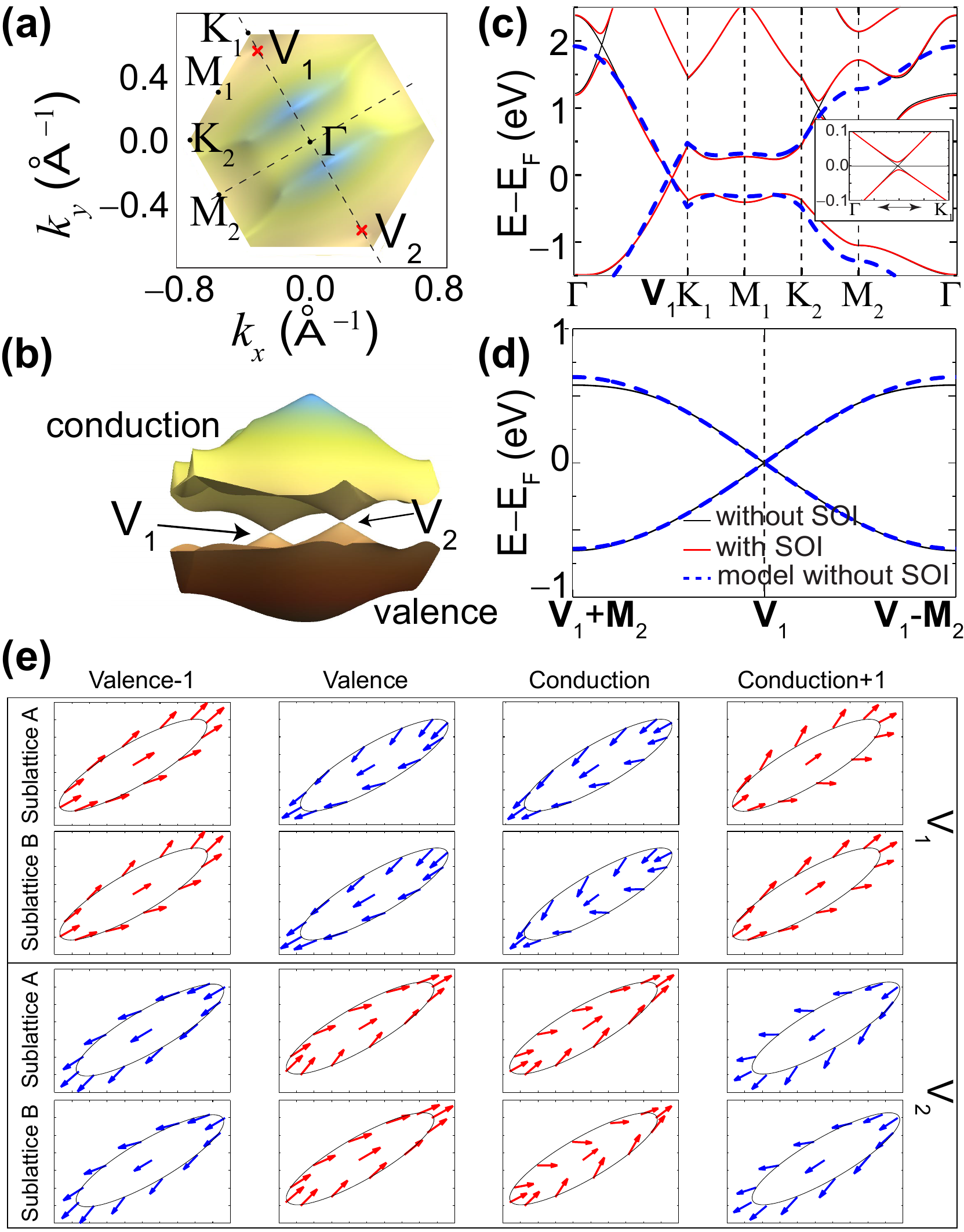}
\caption{(Color online.) (a) Conduction band on the first Brillouin zone, highlighting high-symmetry points and locations of valleys {\bf V}$_1$ and
{\bf V}$_2$ away from the K- and K'-points. (b) The two valleys on the Brillouin zone arise from the two-fold symmetry of the atomic structure. (c-d)
Band structures along high-symmetry lines, including a two-band tight-binding fit. (e) Spin texture resolved over valley ($\boldsymbol{\tau}$), energy,
and sublattice ($\boldsymbol{\sigma}$) degrees of freedom. (The spin projection onto the $z-$axis is of the order of 1\% at most.)}\label{fig:F4}
\end{figure}

 The first Brillouin zone in Fig.~5(a) shows a top view of the conduction band and the high-symmetry points in momentum space. As seen in Fig.~5(b), the
 arrangement of parallel 1D Sn wires gives rise to an electronic structure with only two anisotropic Dirac cones on the First Brillouin zone located away
 from the K-points at positions {\bf V}$_1$ and {\bf V}$_2$ $=\pm$0.85{\bf K}$_{1}$, respectively. From now on we identify the $x-$axis with the line
 joining tin atoms across fluorine bridges. The Fermi velocity is close in magnitude to that of graphene and it is anisotropic: $v_{Fy}=5.4\times 10^{5}$
 m/s [Fig.~5(c)], and $v_{Fx}=2.1\times 10^{5}$ m/s [Fig.~5(d)] and a $2\Delta= 0.02$ eV gap opens due to SOI,\; five times larger
 than the intrinsic gap due to SOI in graphene \cite{Huertas-Hernando}. Phase {\bf 6} transitions from a topological insulator to a trivial
 insulator \cite{stanene}, but the electronic structure of the optimal phase remains robust under larger isotropic strain.

 The electronic dispersion in Fig.~5(b-d) can be understood in terms of a $2\times 2$ $\pi-$electron tight-binding Hamiltonian \cite{Gilles} in
 which an effective coupling $t'$ is set among the tin atoms originally linked by fluorine bridges [thin bonds on Fig.~4(b)],
 and $t$ is the coupling among actual Sn-Sn atoms [thick bonds on Fig.~4(b)]. Using interatomic distances among Sn atoms from Table I we obtain
 the blue dashed lines in Fig.~5(c,d) with $t=0.8$ eV and $t'=\frac{v_{Fx}}{v_{Fy}}t$ which reproduce first-principles results.

  To account for SOI, we realize an oblate low-energy Dirac-Hamiltonian at the vicinity of the {\bf V}$_{1,2}$ points. The relevant subspace is
  four-dimensional at any given valley, and the task is to reproduce the spin texture displayed in Fig.~5(e) where spin projects onto the $+x$ or the
  $-x$ directions while leaving the sublattice (pseudospin) degree of freedom unpolarized. The numerical results on Fig.~5(e) are consistent with a coupling $\tau_z\sigma_x s_x$. Indeed,
  eigenvectors of $\tau_z\sigma_x s_x$ in Table II project spins onto the $-x$, $+x$, $+x$, $-x$ axis parallel to the Sn-F bonds,
   inverting signs at each valley and lacking sublattice polarization, consistently with {\em ab-initio} data [Fig.~5(e)].
    Thus, the low-energy dynamics is given by:
\begin{equation*}
 H=-i\hbar\Psi^{\dagger}(v_{Fx}\tau_z\sigma_x\partial_x + v_{Fy}\sigma_y\partial_y)\Psi + \Psi^{\dagger}(\Delta \tau_z\sigma_x s_x)\Psi.
\end{equation*}
An unprecedented specific coupling of momentum --including direction-- with spin oriented along $\hat{\mathbf{x}}$
and valley degrees of freedom is thus realized by the second term in previous equation. The valley degree of freedom can be addressed
by a bias along the $\mathbf{V}_1-\mathbf{V}_2$ axis that breaks inversion symmetry. Similarly,
a magnetic field along the $\hat{\mathbf{x}}$ axis will break time-reversal symmetry, locking the valley and crystal momentum direction at
the $\mathbf{V}_1$, $\mathbf{V}_2$ points. The dynamics invites the use of 2D fluorinated tin for valleytronic applications.

 We demonstrated the structural stability of HB tin and HB lead and discussed their electronic properties, showed that tin and lead are not
 viable routes towards topological fullerenes, and discovered the structural, valley, sublattice and spin properties of optimal fluorinated stanene. We are
 grateful to G. Montambaux, M. Kindermann and L. Bellaiche, and acknowledge the Arkansas Biosciences
 Institute (P.R. and S.B.L.); the Faculty Development and Research Committee, grant OSPR 140269 and the FCSM Fisher General Endowment at Towson
 University (J.A.Y.); the Spanish MICINN,  grant FIS2012-34858, and European Commission FP7 ITN ``MOLESCO,'' grant number 606728 (V.M.G.S. and J.F.); and
 a Ram\'on y Cajal fellowship RYC-2010-06053 (V.M.G.S.). Computations were carried out at
 Arkansas and TACC (XSEDE TG-PHY090002).


\begin{thebibliography}{35}
\bibitem[{not({\natexlab{a}})}]{note1}
\bibinfo{note}{Phonon dispersion calculations were performed with the Vibra
  tool. The atomic displacements used in computing force constants were as
  small as 0.01 \AA, we set the electronic tolerance smaller than 10$^{-5}$ eV,
  and the mesh cutoff for the real-space grid was larger than 800 Ry. Large
  supercells were needed in order to eliminate spurious negative frequencies,
  and a $11\times 11\times 1$ ($13\times 13\times 1$) supercell containing 242
  (338) atoms for the calculations of high-buckled tin (lead). Due to a heavier
  computational cost, the calculations that included SOI were carried out on
  $9\times 9\times 1$ supercells, containing 162 atoms. A $k-$point sampling of
  at least $6\times 6\times 1$ in all phonon calculations ensures a properly
  converged electronic density.}
\end{thebibliography}
\end{document}